# Automated Line Tracking of λ-DNA for Single-Molecule Imaging

Juan Guan,[a] Bo Wang,[a] and Steve Granick[a,b,c]

Departments of Materials Science and Engineering,[a] Chemistry,[b] and Physics[c]

University of Illinois, Urbana, IL  61801

## Abstract

We describe a straightforward, automated line tracking method to visualize within optical resolution the contour of linear macromolecules as they rearrange shape as a function of time by Brownian diffusion and under external fields such as electrophoresis.  Three sequential stages of analysis underpin this method: first, "feature finding" to discriminate signal from noise; second, "line tracking" to approximate those shapes as lines; third, "temporal consistency check" to discriminate reasonable from unreasonable fitted conformations in the time domain.  The automated nature of this data analysis makes it straightforward to accumulate vast quantities of data while excluding the unreliable parts of it.  We implement the analysis on fluorescence images of λ-DNA molecules in agarose gel to demonstrate its capability to produce large datasets for subsequent statistical analysis.

**Introduction**

This paper focuses on methods of image analysis that enable one to go beyond tracking the position of single molecules, to also analyze their internal conformations, in instances when molecules display shape fluctuations under various conditions ranging from thermal equilibrium to deformation under mechanical stress or other external field. As these changes are rapid, we are interested in real-time measurements during which the need to acquire data rapidly without signal averaging introduces experimental uncertainty. These problems of tracking internal degrees of freedom, which become technically feasible when the size of macromolecules exceeds the resolution of a microscope in one or more directions, apply especially to tracking biological macromolecules, among them filaments such as actin [1,2] and more flexible molecules such as DNA [3-11].

Methods of image analysis could not be applied until recently to problems of this kind; indeed, in the early days the data were typically acquired by video microscopy, as for example in measurements of DNA and actin using fluorescence microscopy [1-12]. Almost from the beginning, special attention was given to direct observation of polymer conformations perturbed from equilibrium by mechanical force or electric field, but quantification was held back in part by the limited resolution of video cameras, in part by the inability of routinely analyze images using methods that would require significant computing power and data storage capacity. It is understandable that quantification to date has concerned largely radius of gyration and coarse measurement of shape anisotropy by measures such as the long axis and short axis components of fluorescence images in two-dimensional projection in the plane of a microscope [13,14].

In addition, many of the early studies suffer from few statistics and involve analyzing a small number of images [4,11,12]. Yet from the beginning of this line of research, it was evident [4-10] that single-molecule analysis of chain conformations holds the promise to measure the distribution of chain conformations whose averages enter into important ensemble-averaged quantities, such as rheology and electrophoretic mobility. Nowadays, it is feasible to use inexpensive personal computers to facilitate analysis of polymer conformations with large statistics and high accuracy. This study is in the spirit of an earlier pioneering automated line tracking method, introduced to analyze actin [15], that works well to analyze filamentous macromolecules that are stiff. The image analysis methods described below are designed to apply equally to more flexible macromolecules such as DNA.

The plan of this paper is to dwell primarily on the methods of automated image analysis we have developed to deal with this problem and to explain the logic that prompted the choice of those methods rather than others. These methods consist of three elementary stages: first, to discriminate the shapes of macromolecules from noise, which we refer to as "feature finding"; second, to approximate those shapes as lines, which we refer to as "line tracking"; third, to discriminate reasonable from unreasonable fitted conformations in the time domain, which we refer to as "temporal consistency check." Finally, this paper presents two examples of applying the method. Although the methods are general and should readily be adaptable to other visualization techniques, here we illustrate the arguments using sample images from DNA in a fluorescent microscope. We focus on DNA that is driven by electric fields to adopt relatively extended conformations and a variety of shapes.

These methods, which build upon image processing methods reported earlier from this laboratory which at that time did not consider internal degrees of freedom in the tracking of single molecules [16,17], amount to approximating the true shape of linear macromolecules by renormalized curved lines, lines that are optimized to describe the shape within the limits of optical resolution. As methods of super-resolution imaging [18] gain more widespread use, one can expect even more need for image analysis methods to analyze internal chain conformations.

**Experimental Methods**

The image analysis presented below was performed on data acquired in the following manner.

Fluorescence microscopy. Data were acquired in epifluorescence mode, typically at a frame rate of 33 fps. A 532 nm excitation laser was focused at the rear focal point of an oil immersion objective (Zeiss, α-Plan Fluor 100×, NA = 1.45) with 2.5x post-magnification to image with a resolution of 64 nm x 64 nm per pixel. Fluorescence images were collected through the same objective and detected by a back-illuminated electron multiplying charge-coupled device (EMCCD) camera (Andor iXon DV-897 BV) after filtering out light from the excitation laser. The movies were converted into digital format and analyzed. A typical dataset consists of 30 movies, each of them consisting of 4,000 frames per movie acquired at 33 fps. The resulting dataset of conformations typically amounts to $>10^4$ from thousands of molecules.

DNA samples. Lambda-DNA (48.5 kbp, Promega) was labeled by covalently attaching dye, a RhB derivative (Mirus) to heteroatoms on DNA, at a labeling density of roughly one dye per 5 base pairs. Single-molecule measurements of DNA chain conformations were made in a miniature gel electrophoresis setup using agarose gel (final concentration 1.5 % (w/v)), in 0.5x TBE buffer (45 mM Tris, 45 mM borate, 1mM EDTA), the DNA being at picomolar concentration. Anti-photobleaching agent, ascorbic acid (SigmaAldrich), was present at a final concentration of 10 mM. A DC voltage was applied across two Pt electrodes to generate an electric field ranging from 6 to 16 V/cm.

## Methods of Image Analysis

The purpose here is to quantify, within the limits of optical resolution, the linear shapes of CCD images that are noisy, faint, and diffraction-blurred. In the methods described below, we take the approach that it is better to reject data from consideration than to improperly include it in subsequent analysis. Therefore, especially in the third step of analysis consisting of checks of data consistency in the time domain, we reject up to 90% of the data acquisition frames. Examples of the variety of raw data are given in Fig. 1, panels c, d, and e.

*Identifying each DNA molecule.* For various practical reasons, despite one's most careful efforts to optimize image quality, the signal-to-noise ratio is never satisfactory in unprocessed images (Fig. 1a). First, the finite DNA concentration results in background fluorescence from other molecules that contribute to the image in spite of being located outside the focal plane. Secondly, the DNA that we track moves rapidly under electrophoresis, leaving limited time to collect photons at each position. Third, the elongation of chains under electric field lessens the local fluorophore density. We find that applying a Gaussian filter locally at each pixel, with Gaussian weighted contribution of intensity from neighboring pixels, significantly reduces noise without compromising the main features of a DNA molecule (Fig 1 b).

As each image consists primarily of background, only a few pixels representing the dilute DNA molecules of interest, we estimate background intensity noise from the mean intensity at each pixel, and estimate noise level from the variance of these pixel intensities. After subtracting this background from each pixel in the image, we retain for subsequent analysis only pixels at which the residual intensity exceeds a threshold, typically set to be several times the noise level. Control checks show that subsequent line tracking does not depend sensitively on the detailed choice of threshold value, nor on whether the image of the DNA is included in estimation of the background.

Next, the image analysis makes judgments to decide whether two features close in space belong to a single molecule. It is a problematical question because low signal-to-noise ratio pixels may in principle register the absence of DNA, but also in principle may indicate that low intensity parts of the molecule, in some cases even vanishingly-low intensity due to various reasons discussed earlier in this section. Fig 2a illustrates a stretched DNA whose middle part presented an intensity less than an intensity threshold. To analyze this situation, first we connect pixels whose distance is less than a threshold and consider tentatively the connected pixels to come from a single molecule. Depending on the threshold selected, the pixels in an image then cluster into either a single grouping, or several. For example, using a threshold distance of 8 pixels, Fig 2b presents two clusters of pixels, whereas a threshold of 9 pixels implies just a single cluster of pixels; the difference depends on whether critical connections are allowed or disallowed by that threshold (white lines in Fig. 2b). To automate the process of line tracking, in order to perform this grouping of pixels first we apply a fixed threshold and analyze the best line through the implied grouping. Then unreasonable lines are removed at a later stage of analysis through the automated temporal consistency checks discussed in a later section.

*Tracing a line through each DNA molecule.* This relies on the concept of a minimum spanning tree, a concept in graph theory which quantifies the shortest path length between nodes in an image [19].  Here, the nodes in the image are the pixels of the molecule defined from the previous section.  Fig. 3a illustrates the concept schematically and Fig 3b illustrates it for a given set of data acquired in our experiments.  First, a tree is constructed by connecting every two pixels.  However, since some pixels are bright and others are dim, and we wish to preferentially include pixels of high intensity to minimize the chance of resulting lines being trapped on noise pixels, the connection length between pixels is assigned not simply as the spatial distance by which they are separated, but weighted by the inverse of their sum intensity.  A minimum spanning tree is then generated choosing from existing connections based on each connection length.  We find that this feature of intensity weighing is often necessary but that the exact intensity weighing method is not critical in line tracking;  for example, an alternative weighting, exponential weighting according to intensity, gives similar results.  Searching through this minimum spanning tree, one can find one path between two termini that contains the largest number of pixels:  this is the longest path through this molecule (Fig 3c).

To yield the final line tracking result (Fig 4a), we follow a four-step progressive refinement of the result.  First, to smoothen the line, we perform a polynomial fit, typically a quadratic fit, locally on adjacent pixels (Fig 4b).  The local fit is preferable to fitting the entire line with a polynomial because the overall shape is sometimes highly curved.  An additional advantage of local fitting is that to improve accuracy, one can select at each point which coordinate, horizontal x or vertical y, to fix or fit.  After this process, each pixel is reassigned a new point that has a modified x or y position.

To further smoothen the line, the second step is to fit the cross-section intensity profile at each point along it to a Gaussian intensity profile, the principle being that diffraction-limited images are expected to be described by this function.  The position of each pixel along the line is then accordingly adjusted to the center of the Gaussian (Fig 4c); typically, this adjustment is on the order of one pixel.  In this fitting, the orientation of the cross section is taken to be perpendicular to the tangent line at each point.  As the coordinates at this stage of the analysis are typically not located on integer pixel positions, the intensity at each point is calculated based on the surrounding four pixels (integer coordinates) via linear interpolation.

Third, the minimum spanning tree is recalculated from the points which resulted from the second step (Fig 4d).  The process to do so is the same as before, except that adjacent points are linked without intensity weighing.

Last, high-frequency noise is removed from the line using a wavelet filter (Fig 4e).  It is convenient to employ a discrete wavelet transform at level 2 using Daubechies-16 wavelets.  Then the coefficients are soft thresholded using a universal threshold and the smoothened line is obtained by inverse transforming the coefficients after thresholding.

*Temporal consistency checks.*  The idea is to err on the side of caution:  we aim to exclude questionable data from the dataset that we construct for subsequent analysis.  As the automated nature of this data analysis makes it straightforward to accumulate vast quantities of data, there is no disadvantage to excluding from analysis the unreliable parts of it.

As a premise, we take the view that when image acquisition is rapid relative to those conformational fluctuations of DNA that occur on distances resolvable within optical resolution, a true line is likely to be similar to those close to it in time, but an unreasonable line is unlikely to satisfy this criterion.  Dissimilar lines may arise for a number of uninteresting reasons:  mistaken grouping of data in the first level of analysis to identify the starting DNA molecule;  molecules whose extension is so limited that it hardly exceeds the optical resolution; molecules out of focus; and an error in generating the minimum spanning tree. Reasoning from this premise, we compare each line, acquired at a given moment in time, to its antecedents and progeny over the span of a few seconds. When two lines are similar, both are regarded valid and included in the final set of lines.

To implement this idea, the line tracing the contour of each DNA molecule is divided into 21 notional fiducial markers of equal spacing along the line and the average distance by which a line is displaced from another line within the temporal vicinity of a few seconds is measured against a threshold value, as drawn schematically in Fig 5a. A small average distance means two lines are similar and are likely to both be valid. After this selection, anomalous lines are removed, as illustrated in Figs. 5b and 5c. In practice, we choose the low threshold value of 0.1 μm, less than the diffraction limit, in order to insist on high accuracy.

We find that accuracy of the final set of lines is improved by performing a series of pre-screenings before the selection. This automated process searches for many anomalous features; when any of them is identified, that frame of the dataset is excluded from analysis. First, we remove regions where line length fluctuation with time is frequent and unreasonably drastic. Second, lines that are too short relative to their neighbors in time are considered to be the likely result of partial features due to incorrect grouping of molecules or else to low intensity parts not identified as signal; as these anomalously short lines can potentially introduce large errors in further quantitative analysis, they are removed. Third, we combine intensity and shape anisotropy information to exclude from consideration molecules that are out of focus. Fourth, the size of a candidate molecule is used to exclude noise and partial features. Fifth, the end-to-end distance of a line is considered, relative to the length of that line; if the ends appear to loop together too closely, as can happen when the image area is close to the optical diffraction limit, this also is considered likely to be anomalous, and is excluded from further analysis.

It is true that in principle, the exclusion of data might risk biasing the dataset. Checking the radius of gyration before and after this selection, we find no bias towards a subpopulation of molecules. The distribution of radius of gyration remains unchanged indicating that the criteria, by which lines are considered to be unreasonable, do not depend on size of the molecule. The exception is that the smallest 10% of molecules (radius of gyration smaller than the diffraction limit, which corresponds to less than half the mean radius of gyration of λ-DNA) do not contribute to the final set of lines. Line tracking should not be expected to work well in this situation. In addition, features so small are likely to be either noise or fragmented molecules, and in this respect it is proper to exclude them. Testing directly for hypothetical bias of the data, we also measured average DNA velocity along the electric field direction, with and without the selection of data just described, and found no meaningful difference.

**Application to Specific Systems**

To illustrate how automated line tracking allows visualization of polymer conformation changes in dynamic processes, we now present two examples, both of which will be fully explored in subsequent reports from this laboratory. The point of these examples is simply to illustrate the quality of data that can be obtained routinely, with large statistics and high fidelity, using the image analysis methods presented in this paper.

Fig. 6 illustrates experiments in which a λ-DNA molecule, attached at one end to the agarose gel in which it is embedded, is subjected to repetitive stretch and release by applying a periodic electric field.  In Fig. 6a, which overlays lines tracked from hundreds of frames when this molecule was subjected to repeated stretching and releasing from a square wave, one notices that stretch-retraction events differ subtly from one another, differing not just in times for these processes to be accomplished, but also in the paths by which the molecule is threaded through the gel.  Fig. 6b illustrates a small number of the accompanying length fluctuations during these stretch and release events.  Whereas stretching and recoiling both transpire on the timescale of one second, a recoil process is also clearly evident between two stretches, before the molecule takes on a different path.

Fig. 7 illustrates data from a different experiment:  $\lambda$-DNA migrating through agarose gel under a DC field of 16 V/cm.  This set of data, consisting of $\approx 4 \times 10^4$ evaluations of the length of the molecule, shows clearly the statistical nature of the chain length:  while the mean chain length is well-defined, and the most likely chain length is well defined, the distribution around averages is large.

**Prospects**

In this computer age, with large computing power and digital storage capacity readily accessible, one can use inexpensive personal computers to facilitate image analysis of optical images.  Here we have introduced automated line tracking methods applicable to tracing the linear coarse-grained shapes of biomolecules, those shapes larger than the optical diffraction limit, and have illustrated their application to analyzing the conformations of fluorescent-labeled $\lambda$-DNA when it is stretched by electric fields.  These automated methods enable the facile acquisition of large datasets and by rational extension should readily be adaptable to analysis of data obtained from other visualization methods. It is different in spirit from principal component analysis of DNA, which expresses dynamic information in a virtual phase space of orthogonal basis sets that can be problematical to interpret physically [20].

While the fidelity of tracking reported here is believed to have been optimized within the limits achievable using optical resolution, it is certainly the case that this image analysis is limited in resolution. The line tracking introduced in this study represents a coarse-grained representation of the actual contour of λ-DNA; this is why, for example, even the longest lengths plotted in Fig. 7 are a factor of 3 small than the known contour length of the molecule, 16 μm. Thus, while these methods are well adapted for quantifying time scales of dynamic processes (illustrated in Fig. 6) and also their distributions (illustrated in Fig. 7), the numerical values of the lines do not, at the present time, have one-to-one correspondence with the actual molecular makeup, but should be viewed instead as coarse-grained representations. We remark that this coarse-grained can carry physical meanings, especially when the optical resolution limit coincides with the fundamental length scale of the system; for example, the pore size of agarose gel, ~200 nm, nearly coincides with the diffraction limit in this study. Provocatively, the same also holds for many other biomacromolecular solutions.

**Acknowledgements.** We thank Kejia Chen for discussions. This work was supported by the U.S. Department of Energy, Division of Materials Science, under Award No. DEFG02-02ER46019. For instrumentation, we acknowledge support from NSF-CBET-0853737.

**Figure Captions**

<u>Fig 1</u>. Examples of raw data in which λ-DNA (contour length 16 μm) displays various conformations in an optical microscope. Color bar corresponds to relative intensity within each image. (a) A typical unprocessed image as input raw data. (b) A local Gaussian filter reduces noise significantly. (c),(d),(e) Additional examples of various polymer conformations with final line tracking results (black line) overlaid. Scale bar: 1 μm.

<u>Fig 2</u>. Feature finding: grouping pixels into one DNA molecule or several according to the threshold distance between pixels. (a) An example of an image in which only a fraction of the middle pixels have high enough intensity to be recognized as signals. (b) The connections between pixels discriminate whether they cluster into one grouping or several. The leftmost two white connections are critical – if not present, this image is treated as two groupings, each one from a different molecule. The pink connections are separated by 5-8 pixels. The white connections are separated by 8-9 pixels. Scale bar: 1 μm. Color bar is the same as in Fig. 1.

<u>Fig 3</u>. Line tracking: using minimum spanning tree analysis to find a longest path through pixels to identify a line through the DNA molecule. (a) Schematic illustration of the notion of minimum spanning tree. The edges belonging to the minimum spanning tree are drawn in solid lines whereas other edges are drawn with dashed lines. Next to each edge, the indicated numbers specify the relative weight. (b) Example of an intensity-weighted minimum spanning tree (white) overlaid onto real data. (c) For the data in panel (b), a longest path (thick white line) is identified from analyzing the minimum spanning tree. Scale bar: 1 μm.

<u>Fig 4</u>. Four-step refinement of the line identified in each single image. The highlighted region in panel a, magnified in panels b-e, is used to illustrate how a line changes from each step to the next step during the refinement procedure. In panels b-e, starting lines and symbols are white, and the line resulting from that step is drawn in black. (a) Starting from a longest path connecting pixels (white, same as in Fig 3c), a four-step refinement yields the final line tracking result (black). (b) A polynomial fit. (c) A Gaussian fit to re-center. (d) A minimum spanning tree through the resulting points in c. (e) Wavelet smoothening. See text for details.

Fig 5. Tests of temporal consistency: selecting reasonable lines from temporal comparison of images at different times. (a) Each line is overlaid with 21 notional fiducial points. Within the temporal vicinity of a few seconds, the average distance between each point on any two lines is evaluated (red lines). When this is less than a threshold designation of similarity, the lines are deemed valid and kept. Lines that fail this similarity criterion are rejected as physically unreasonable. In this typical example, the threshold is 0.1 μm, less than the diffraction limit. (b) Raw data: four lines from the level of analysis in Fig. 4. (c) Outcome of the test for temporal consistency: line 1 is rejected. The text describes additional pre-screenings included in these checks.

Fig 6. An example of applying automated line tracking analysis. A λ-DNA molecule with one end attached to the agarose gel in which it is embedded is subjected to repeated stretch and release using a square wave electric field alternating with period about 10 sec between 16 and 0 V/cm. (a) Overlay of all lines tracked over the time of 25 seconds, at 33 frames per second, showing the distribution of paths by which the molecule threads through the gel. Color bar denotes time. Scale bar: 1 μm. (b) Lower panel: Square-wave electric field applied to the DNA over 25 sec. Upper panel: Length fluctuation of lines tracked during this time window, showing stretch and retraction. Vertical dashed lines show the time at which the electric field switches on and off.

Fig 7. A second example of applying automated line tracking analysis. This figure shows a histogram of length defined by line tracking, as λ-DNA migrates through agarose gel under a DC electric field of 16 V/cm. The dataset was 30 movies, each of them consisting of 4,000 frames per movie acquired at 33 fps. Number of observations is plotted against length with bin size 0.2 μm.

Fig. 1

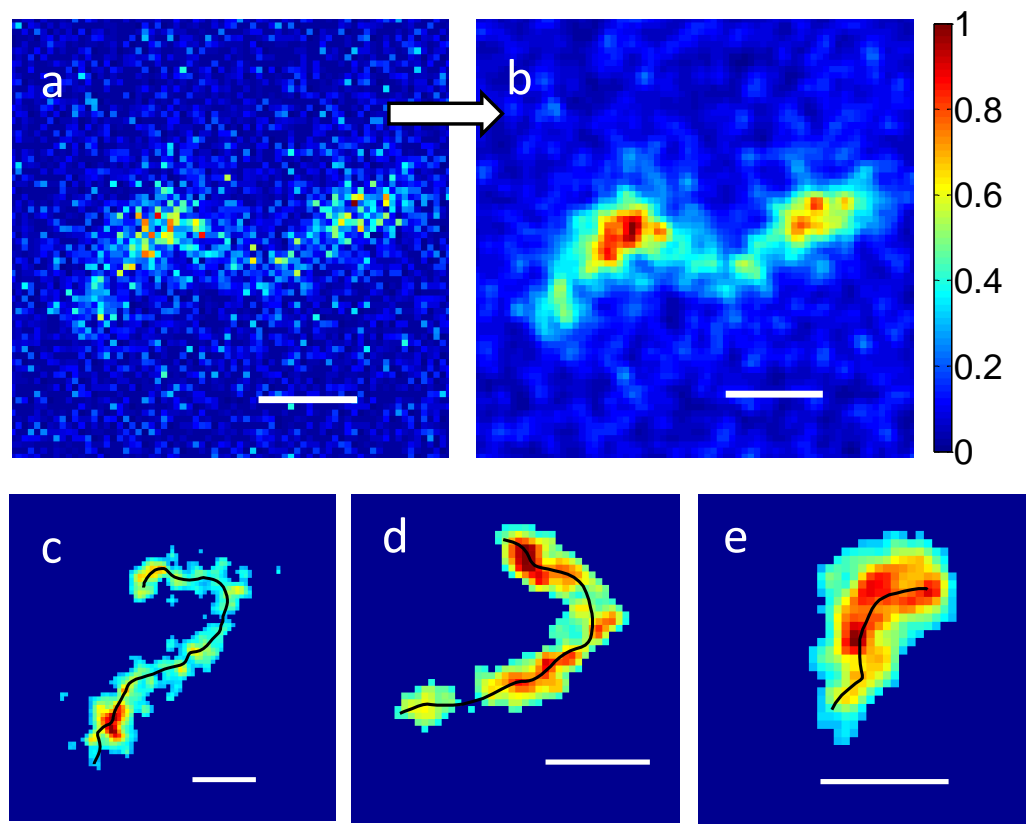

Fig. 2

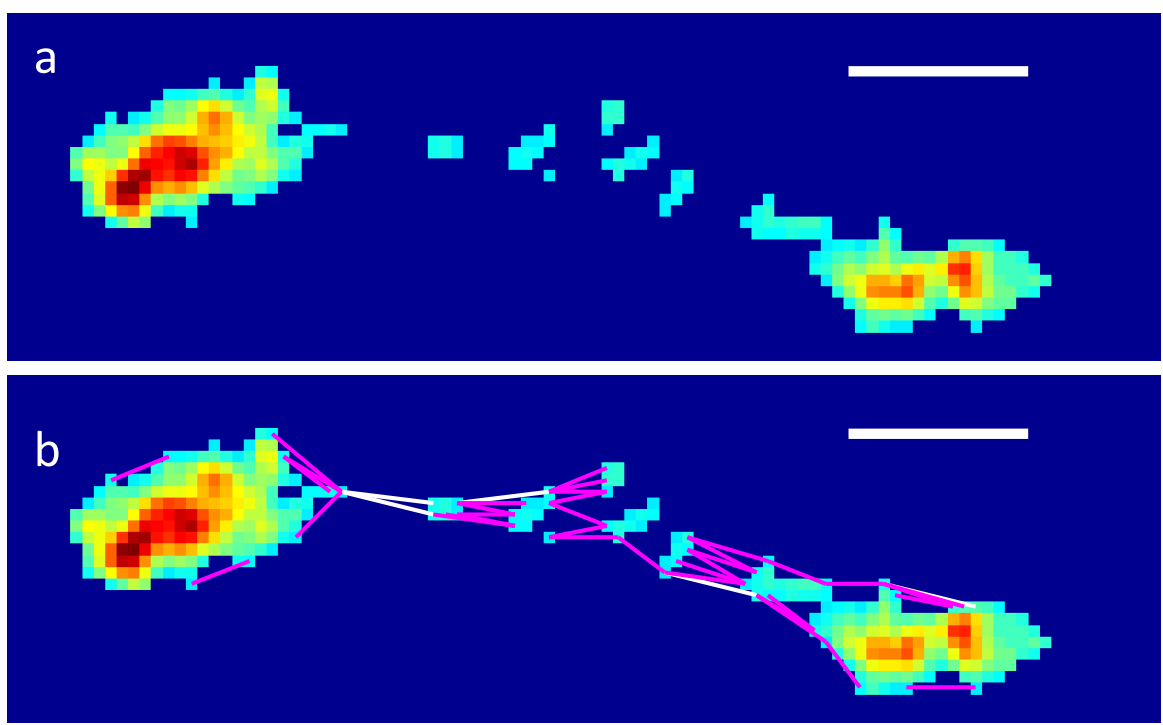

Fig. 3

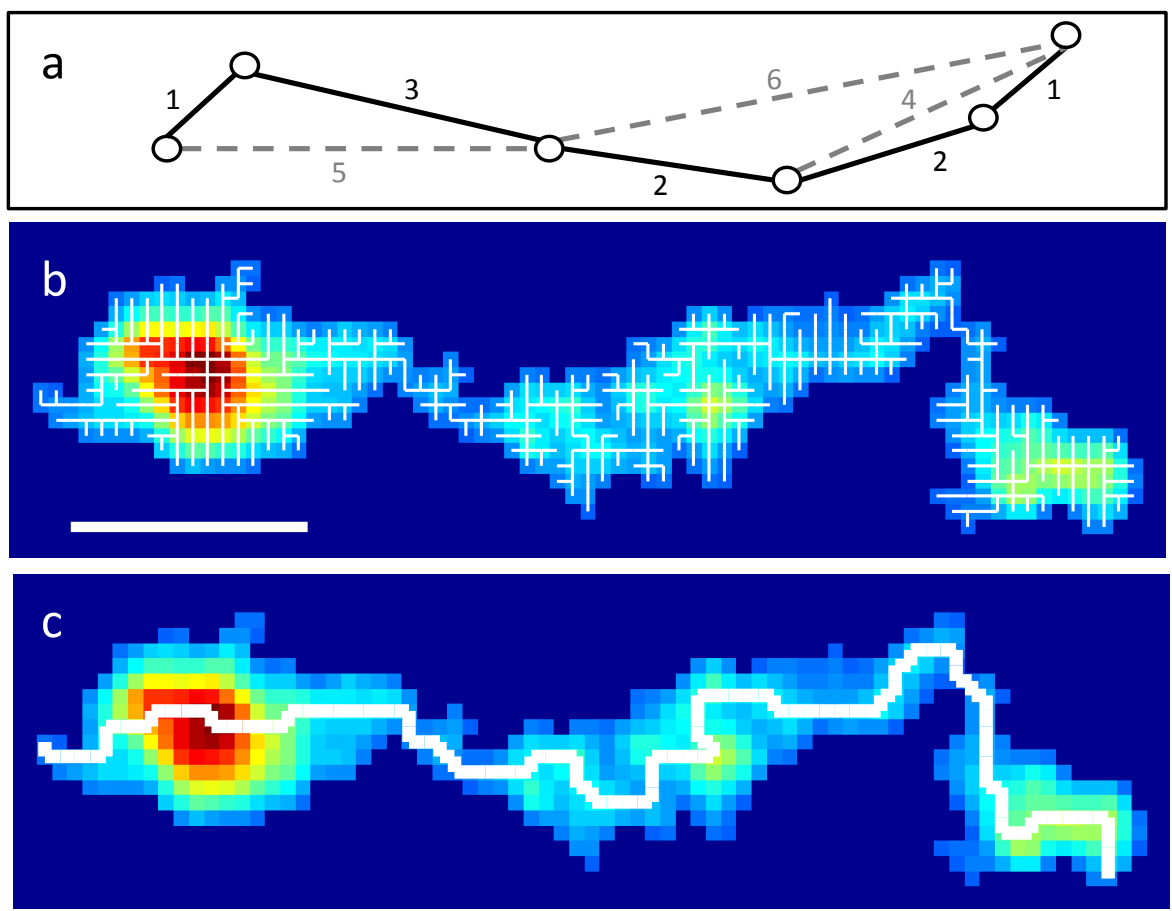

Fig. 4

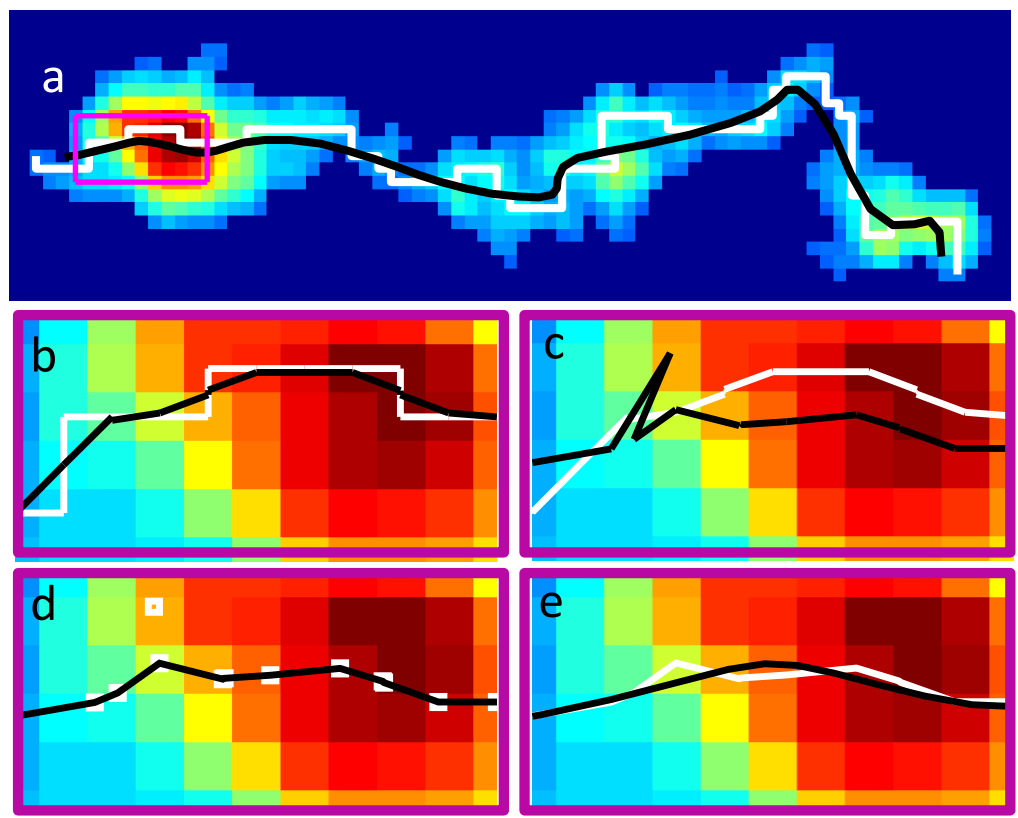

Fig. 5

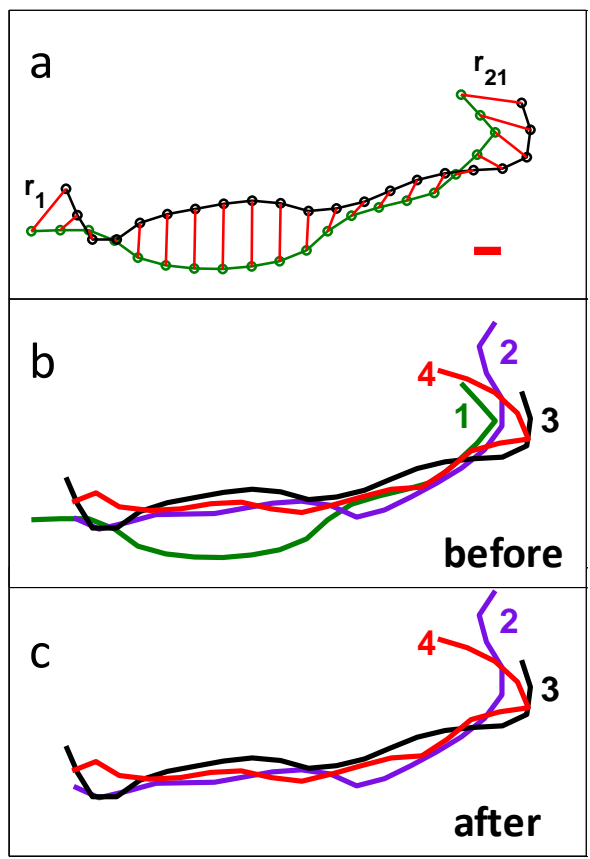

Fig. 6

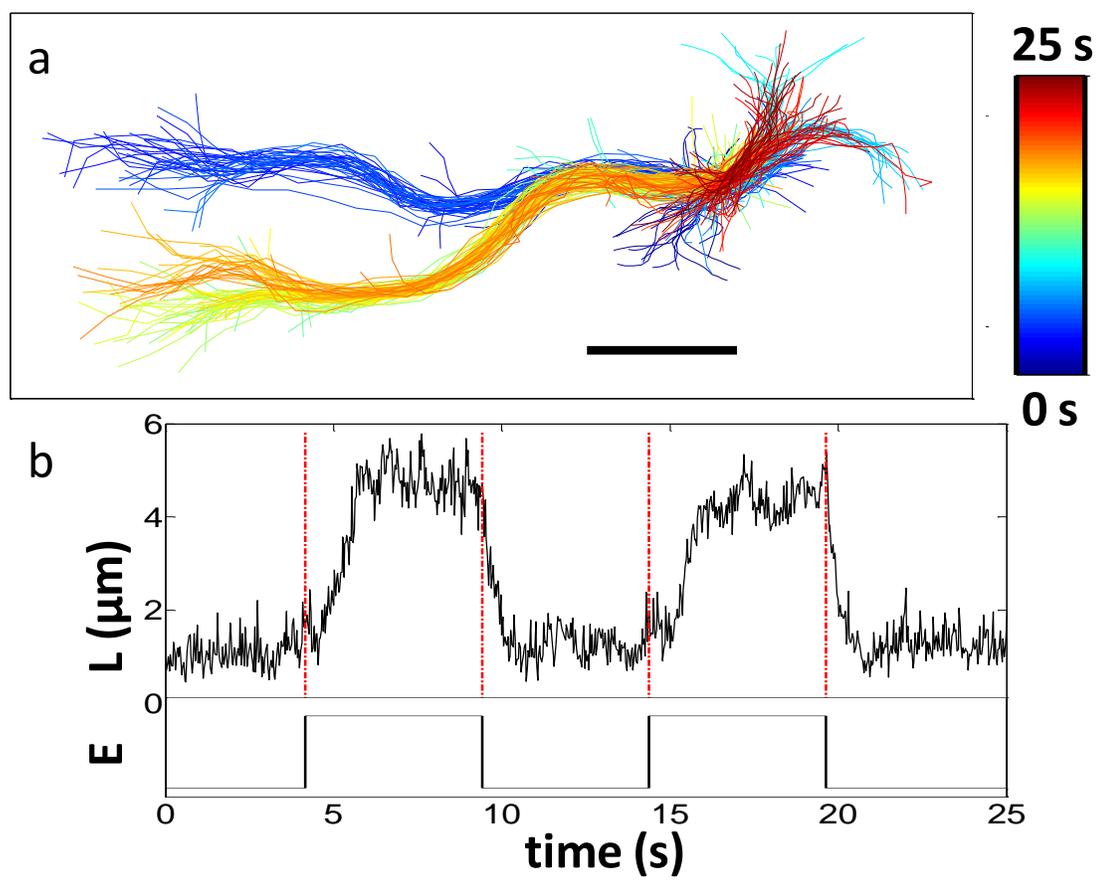

Fig. 7

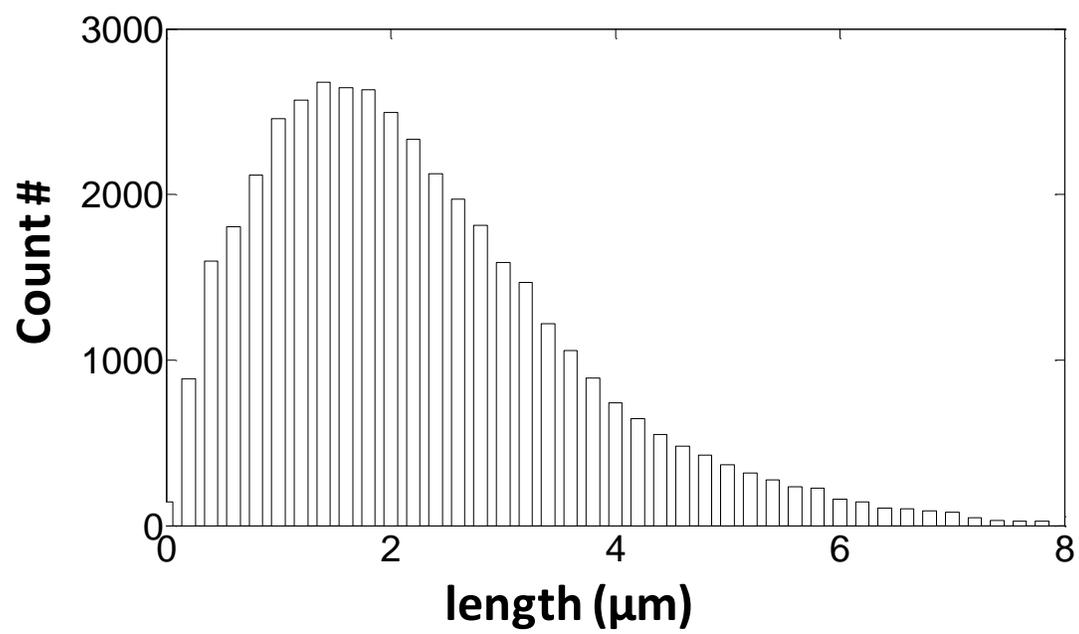